\documentclass[%
 reprint,
 amsmath,amssymb,
 aps,
pre,
]{revtex4-2}

\usepackage{graphicx}% Include figure files
\usepackage{dcolumn}% Align table columns on decimal point
\usepackage{bm}% bold math
\graphicspath{{./Fig/}}

\usepackage{xcolor}

\begin{document}
\preprint{APS/123-QED}

% Main title of the paper
\title{Convergence criteria for self-consistent measures in bipartite networks
}                      

\author{János Török}
\email{torok.janos@ttk.bme.hu}
\affiliation{
Department of Theoretical Physics, Institute of Physics, Budapest University of Technology
and Economics, Műegyetem rkp 3., H-1111 Budapest, Hungary}
\affiliation{MTA-BME Morphodynamics Research Group,
 Budapest University of Technology
and Economics, Műegyetem rkp 3., H-1111 Budapest, Hungary}

\author{Takashi Shimada}
\affiliation{Department of Systems Innovation, Graduate School of Engineering, The University of Tokyo 7-3-1 Hongo, Bunkyo-ku 113-8656 Tokyo Japan}
\affiliation{
Mathematics and Informatics Center, The University of Tokyo, The University of Tokyo 7-3-1 Hongo, Bunkyo-ku 113-8656 Tokyo Japan}

% Third author
\author{Fumiko Ogushi}

% Address/affiliation
\affiliation{Faculty of Mathematical Informatics, Meiji Gakuin University, 1518 Kamikurata-cho, Totsuka-ku, Yokohama 244-8539, Kanagawa, Japan}

\author{Kata Tunyogi}
\affiliation{
Department of Theoretical Physics, Institute of Physics, Budapest University of Technology
and Economics, Műegyetem rkp 3., H-1111 Budapest, Hungary}

\author{János Kertész}
\affiliation{
Department of Network and Data Science, Central European University, Quellenstrasse 51, 1100 Vienna, Austria}

\author{Kimmo Kaski}
\affiliation{Aalto University, School of Science, P.O. Box 11000 (Otakaari 1), FI-00076 AALTO, Finland}

\date{\today}

% Here goes the abstract
\begin{abstract}
Many quantities that characterize %characterizing
network elements are defined in an explicit form and calculated directly from the network structure;  examples of include several centrality measures like degree, closeness, or betweenness. %examples include several centralities (e.g., degree, closeness, betweenness). 
However, there are also implicitly defined quantitative measures, %measures defined in an implicit way, 
which are usually calculated iteratively, in a self-consistent manner, like PageRank or countries' fitness / products' complexity relations. The iteration algorithms involve calculations over the entire %whole
network; therefore, their convergence properties depend on the structure of the network. %structure.
Here, we focus on investigating self-consistently defined quantities in bipartite networks of %defined over 
two sets of nodes where the quantities in one set are determined by the quantities in the other set and vice versa. We derive %introduce %give
an explicit convergence criterion for iterations of these quantities and describe two different approaches %methods
to improve the convergence properties. In the first one, %method,
we identify "problematic nodes" that can be removed or merged, while in the second one %. For the second,
we employ a non-homogeneous regularization scheme that guarantees convergence.
\end{abstract}

\maketitle
\section{Introduction}

The network nodes are characterized by a number of %many
quantities, which depend on the local or global structure of the network and reflect the diversity of applications~\cite{oldham2019consistency}. If their definition is explicit, these quantities can be calculated directly, like in the case of some centrality measures including 
the degree, closeness, and betweenness centrality. However, the quantities 
defined implicitly are often calculated using an iterative self-consistent algorithm. Examples of such measures include 
%Since the works by Hartree and Fock~\cite{hartree1928wave,fock1930naherungsmethode} to solve two quantities of many-body quantum problem iteratively, self-consistent methods have become important in various fields of physical sciences. In case of network science it often features multiple measures for the importance of nodes,
%e.g., degree,  
%Although some of these measures are defined by a specific construction, other intrinsic measures are determined by the structure of the network itself.
eigenvector centrality~\cite{bonacich1972factoring}, PageRank~\cite{page1999pagerank}, SimRank~\cite{jeh2002simrank}, DebtRank~\cite{bardiosa2015DebtRank}, Economic Systemic Risk Index~\cite{diem2022ESRI}, measures of economic complexity~\cite{hidalgo2009building,caldarelli2012network,tacchella2012new,cristelli2013measuring,hidalgo2021economic}, and knowledge complexity~\cite{ogushi2021ecology}.

%For networks the %Such 
The self-consistent methods consist of calculating iteratively the quantity at each node that in turn depends on the values %taken 
at the neighboring nodes, and so on. This process is repeated until convergence is detected. A typical example of such quantitative measure is %the 
eigenvector centrality~\cite{bonacich1972factoring}, in which %where 
the importance $x_i$ of a node $i$ is determined by the importance of its neighbors. In this process the %The 
$n$-th iteration is of %has 
the form
%is a good example of a measure that is defined in a self-consistent manner:
$\mathbf{x}^{(n)} = \mathbf{A} \mathbf{x}^{(n-1)}$, where $\mathbf{A}$ is the adjacency matrix of the network. Here the convergence to a unique fixed point is guaranteed by the Perron-Frobenius theorem---a rigorous result that, for this linear measure, settles once and for all on which networks the quantity is well defined. For the nonlinear self-consistent measures that have since proliferated across the network-science literature no comparable guarantee is available, and little is known in general about how their convergence depends on the network structure and link weights. Beyond its practical relevance, this question is also of theoretical interest: since a measure can be interpreted only on the networks for which it is well defined, a rigorous criterion that links convergence to the network structure helps clarify the scope of these quantities and complements the existing results on their numerical stabilization. %In this case convergence to a unique fixed point of the iteration is assured by the Perron-Frobenius theorem. However, there is little known in general about the dependence of the convergence of self-consistent algorithms on the network structure and the link weights.
In the following, we will focus on investigating this issue for %question in the case of 
bipartite networks.  
%where the existence and convergence to the unique %fixed point $x_*$ is guaranteed as
%the Perron-Frobenius eigenvector of $\mathbf{G}$ with eigenvalue $1$,
%$\vec{x}_* = \mathbf{G} \vec{x}_*$. \textit{SimRank}~\cite{jeh2002simrank} is another good example in which the similarity 

Quantitative characterization of bipartite systems through iterative self-consistent measures has emerged as a powerful approach to uncover hidden structural and dynamical patterns in various domains, ranging from economic complexity to information ecosystems~\cite{straka2018ecology,hidalgo2009building,caldarelli2012network,ogushi2021ecology}. A prime example are the Fitness-Complexity measures that were
originally formulated to evaluate the competitiveness of countries and the complexity of their exported products by leveraging the bipartite network that connects countries and products~\cite{hidalgo2007product,tacchella2012new,cristelli2013measuring}. The nonlinear iterative
scheme of~\cite{tacchella2012new} defines fitness and complexity measures through mutually dependent update rules, capturing the nested and hierarchical structure embedded in the underlying data. Over the past decade these metrics have grown into a broad research program in economic complexity, with applications to economic growth, development, and competitiveness across countries, regions, and firms~\cite{hidalgo2021economic}. The same self-consistent construction has, however, well-documented stability problems: the Fitness and Complexity values evolve non-monotonically and a fraction of them decay towards zero as the iteration proceeds, so that the outcome can depend on the stopping rule~\cite{pugliese2016convergence,angelini2017complex}.

This issue has motivated a substantial body of work on the convergence of the algorithm. Early studies, including~\cite{pugliese2016convergence}, provided important insight by analyzing convergence patterns for specific classes of biadjacency matrices and by proposing heuristic criteria for stability. To stabilize the dynamics, a non-homogeneous (regularized) variant of the algorithm was introduced by Servedio \textit{et al.}~\cite{servedio2018stable}, where an additive constant prevents values from collapsing to zero and, after a suitable rescaling, recovers the original metric in the limit of a vanishing regularizer. The convergence of this regularized map has since been placed on firm theoretical footing: it is equivalent to the classical Sinkhorn--Knopp matrix-scaling iteration~\cite{mazzilli2024equivalence}, and it can be derived as the minimizer of a convex cost function, which guarantees a unique fixed point and global convergence for broad classes of networks~\cite{bellina2026cost}. What remains comparatively less understood, and is arguably the more fundamental question, is an explicit, \emph{a priori} criterion that predicts directly from the network structure whether the original, parameter-free measure is well defined---that is, whether the iteration converges at all---and that pinpoints the structural origin of non-convergence. Whereas regularization restores numerical stability by modifying the dynamics, such a criterion characterizes the measure itself: it delineates the class of networks on which the quantity exists, independently of any algorithmic remedy, and thereby plays the role for these nonlinear measures that the Perron-Frobenius theorem plays for the linear ones. We regard establishing this criterion as a valuable scientific result in its own right, and it---together with the structure-preserving interventions it suggests---is the focus of the present work.

Related iterative constructions have surface in a seemingly distinct context of analysing a digital ecosystem of Wikipedia's network structure, %Concurrently, related iterative constructions have surfaced in seemingly distinct contexts, such as in the analysis of digital ecosystems, as exemplified by the investigation of Wikipedia's network structure, 
in which analogous bipartite interactions between editors and pages exhibit complex coevolutionary dynamics~\cite{ogushi2021ecology}. These studies underscore the broad relevance of having iterative frameworks of self-consistency and motivate finding a unifying theoretical treatment to address rigorously the convergence properties inherent to this class of algorithms. In the present study %this work,
we derive a precise and analytically tractable convergence criterion that is valid
for a wide family of nonlinear self-consistent measures, thereby expanding the theoretical foundation needed to apply these methodologies confidently across disciplines. The criterion is independent of the network weights and depends solely on the network structure; it identifies, from the bipartite topology alone, the configurations responsible for non-convergence. This structural viewpoint is complementary to the regularization approach~\cite{servedio2018stable,mazzilli2024equivalence,bellina2026cost}: rather than modifying the dynamics, it explains \emph{why} convergence fails and thereby informs minimal, structure-preserving interventions. Building on it, we discuss and compare three remedies for non-convergence---node removal, node merging, and regularization---and apply them to the international-trade and Wikipedia bipartite networks.

\section{Measures}

Here we define the two similar self-consistent measures for bipartite networks ~\cite{tacchella2012new,ogushi2021ecology}. Bipartite networks are defined on two sets of nodes, where links connect nodes from different sets and the iteratively defined quantity for nodes of one set depends on the characteristic measure of the nodes of the other set and vice versa. For the trade networks the two sets are the countries and the products they export and the measures are \textit{Complexity} $Q$ for products and \textit{Fitness} $F$, for countries~\cite{tacchella2012new}:
\begin{equation}
\tilde F^{(n)}_e = \sum_\alpha w_{e\alpha} Q^{(n-1)}_\alpha 
\qquad
\tilde Q^{(n)}_\alpha = \left(\sum_e \frac{w_{e\alpha} }{F^{(n-1)}_e }
\right)^{-1},
\end{equation}
where Greek and Latin letters index the 
products and countries, respectively, $n$ denotes the number of recursive iterations, and $w_{e\alpha}$ is the weight of the link connecting the nodes $e$ and $\alpha$.  The values of $F$ and $Q$ are normalized after each
iteration step:
\begin{equation}
F_e^{(n)}
=
\frac{\tilde F_e^{(n)}}{\left\langle \tilde{F}^{(n)} \right\rangle_{cntr.}}
\qquad
Q_\alpha^{(n)}
=
\frac{\tilde Q_\alpha^{(n)}}{\left\langle \tilde{Q}^{(n)} \right\rangle_{prod.}},
\end{equation}
where $\langle \cdot \rangle_{cntr.}$ and $\langle \cdot \rangle_{prod.}$ represents the average over the countries and products, respectively.
The \textit{measures of scatteredness} $D$ and \textit{complexity} $C$ introduced by \cite{ogushi2021ecology} are analogous to the previous construction, with $D\equiv F$ and $C\equiv 1/Q$:
\begin{align}\label{Eq:origCD}
\tilde{D}^{(n)}_e = \sum_\alpha \frac{w_{e\alpha}}{ C^{(n-1)}_\alpha }
\qquad&
\tilde{C}^{(n)}_\alpha = \sum_e \frac{w_{e\alpha} }{D^{(n-1)}_e },
\cr
D_e^{(n)}=\frac{\tilde{D}_e^{(n)}}{\left\langle \tilde{D}^{(n)} \right\rangle_{edt.}}
%C_e^{(n)}=\frac{\tilde C_e^{(n)}}{\frac{1}{N_C}\sum_f \tilde C_f^{(n)}}
\qquad&
C_\alpha^{(n)}
=
\frac{\tilde C_\alpha^{(n)}}{\left\langle \tilde{C}^{(n)} \right\rangle_{art.}},
%D_\alpha^{(n)}=\frac{\tilde D_\alpha^{(n)}}{\frac{1}{N_D}\sum_\beta \tildeD_\beta^{(n)}}
\end{align}
where $\langle \cdot \rangle_{edt.}$ and $\langle \cdot \rangle_{art.}$ denote the average over Wikipedia editors and articles, respectively.
These quantities have been shown to correlate with human-determined properties of the systems, but they suffer from the significant drawback of not always converging. For the trade network the mapping $D\equiv F$, $C\equiv 1/Q$ means that $D_e$ is simply the country Fitness---so that a large $D_e$ marks a competitive, highly diversified country---while $C_\alpha$ is the inverse of the product Complexity. Due to their symmetrical nature, we will use the $C$ and $D$ measures for our calculations, but all the results carry over to the $F$ and $Q$ measures through this equivalence. We stress that the equivalence is exact at the level of the (un-normalized) iterative map and of the resulting ordering of the nodes; the per-step normalization by an arithmetic mean differs between the two parametrizations, since $\left\langle 1/Q\right\rangle\neq 1/\left\langle Q\right\rangle$, but this is an overall rescaling that cancels in the convergence analysis below and therefore does not affect whether the iteration converges.

Self-consistent measures are inherently problematic for disjoint networks, as there is no interaction between components apart from normalization. This 
can continuously deplete a component, driving its
measure values to zero. Therefore, here
we only consider connected graphs. However, even for connected graphs one may find scenarios when measures of certain nodes slowly go to zero, which due to normalization results in a state in which a few nodes will have
a finite measure value, while others go to zero. We will call this state non-convergent, since the relative changes in the values of the measures
do not vanish.

This notion of \emph{value convergence} should be distinguished from the weaker \emph{rank convergence} often used in practice, where the iteration is stopped once the ordering of the nodes stabilizes, even if the underlying values keep drifting~\cite{cristelli2013measuring,pugliese2016convergence}. Since the metrics are ultimately used to rank countries and products, one might expect rank convergence to be sufficient. The split into a vanishing group and a finite group is, however, not a benign numerical detail but a structural property of the network, governed by a sharp topological condition derived below. When the network sits close to this condition, the split is dominated by a few marginal nodes, and the ranking itself becomes sensitive to them---as we demonstrate later for the anomalous top ranking of the Philippines, where the convergence issues studied here shift the ranking in a non-negligible way, even when it appears to have stabilized. We therefore adopt the stricter, value-based criterion throughout.

\begin{figure}
\centering
\includegraphics[width=.5\textwidth]{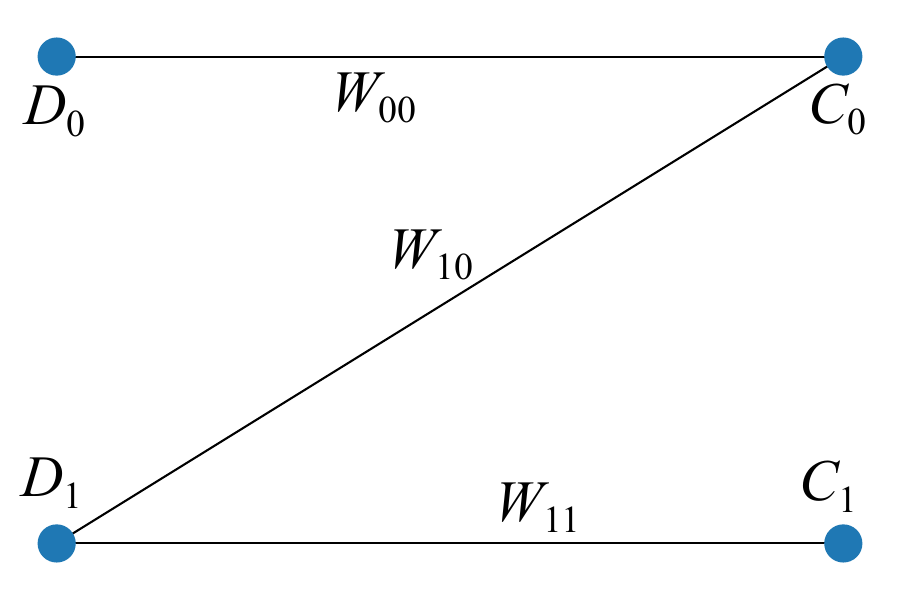}
\caption{The simplest connected bipartite network with at least two nodes in each group.}
\label{Fig:22}
\end{figure}

The mechanism behind such non-convergence is already visible in the simplest connected bipartite network, which has two nodes in each set (Fig.~\ref{Fig:22}). For convenience and since we will primarily use trade data, we refer to the groups as countries (on the left) with property $D_e$ and products (on the right)
with property $C_\alpha$.  For this example, let the link weights be
$w_{00}=w_{10}=w_{11}=1$. From the initial values
$D^{(0)}_0=D^{(0)}_1=C^{(0)}_0=C^{(0)}_1=1$, the system quickly reaches a state where $D_0=C_1\simeq 0$ and $D_1=C_0 \simeq 2$. In fact, the absorbing state $D_0=C_1=0$ is itself a fixed point of the normalized map. Before normalization, $D_1=C_0$ tends to infinity due to the zeros in $D_0, C_1$, while $D_0, C_1$ receives only moderate values from the finite $D_1, C_0$. After normalization, these moderate values will become
zero because of
the infinite values in their respective pairs.  Interestingly, changing the weights does not resolve this issue: the configuration fails to converge for any choice of weights. We note that the two failure modes are distinct here---before normalization the dominant pair $D_1=C_0$ grows without bound (a genuine divergence), whereas after normalization all values stay bounded but never settle, the relative changes failing to vanish; it is this latter, normalized behaviour that we term non-convergence.
This example demonstrates that a measure successfully applied to numerous empirical networks \cite{tacchella2012new,ogushi2021ecology} already fails to converge in the simplest possible case.  Therefore, it is important  %we aim 
to address the question of when convergence occurs and to provide a criterion or an algorithm to resolve this issue. 

\section{Test networks}

\begin{figure}
\centering
\includegraphics[width=.90\columnwidth]{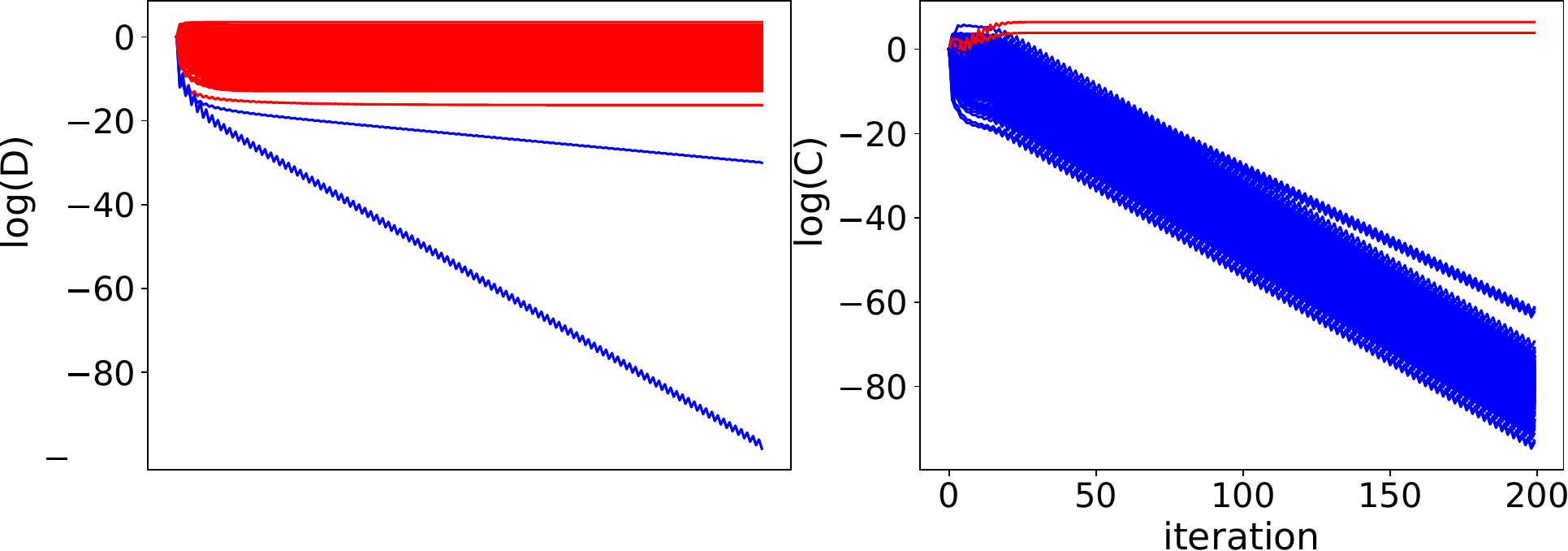}
\includegraphics[width=.90\columnwidth]{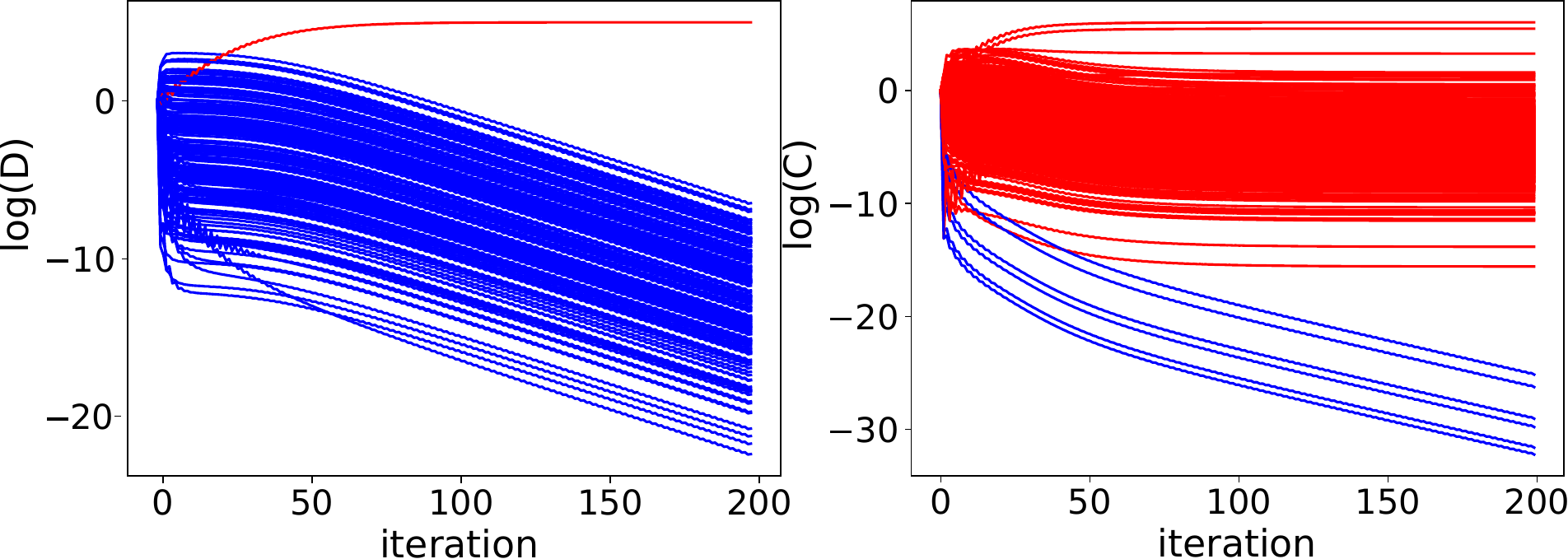}
\includegraphics[width=.90\columnwidth]{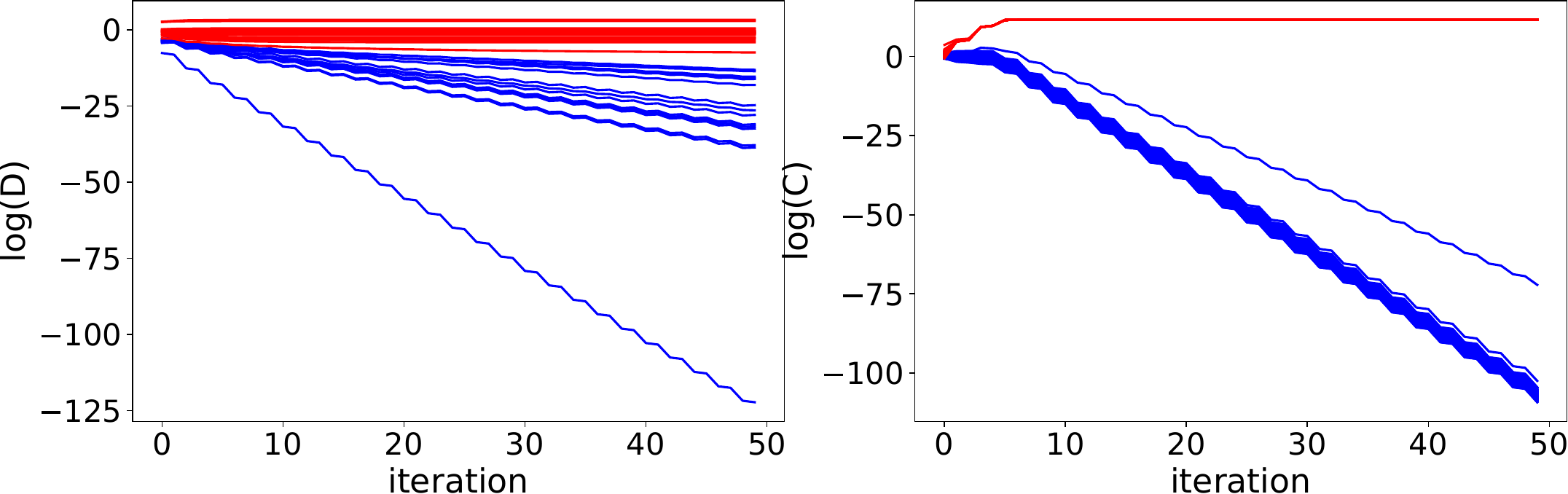}
\caption{Sample evolution of the self-consistent measures $C$ and $D$
during the iteration. The vertical axis shows the logarithm of the
measure and the horizontal axis the iteration number $n$. Left column:
$D$ values; right column: $C$ values. Top row: trade network, 1965
export data; middle row: trade network, 2015 export data; bottom row:
Wikipedia editor--article network (top 942 editors and articles edited
by them; only a selected subset of curves is shown).
Each line follows one node: a $D$ value for a country (trade) or an
editor (Wikipedia), and a $C$ value for a good (trade) or an article
(Wikipedia). Red curves mark measures that remain finite after
normalization, while blue curves decrease to zero. }
\label{Fig:samplenoconv}
\end{figure}

The convergence problem is more prevalent when real data is analyzed.
We chose two datasets for illustration purposes. It is not the aim of
this paper to draw conclusions from the behaviour of the measures.

We used two data sources: (i) the export trade data from the
publication~\cite{hidalgo2007product}. Countries were determined by the
three-letter country codes, products by their id in the data. A
bipartite weight matrix was constructed using the value of the trade.
Data points with N/A or zero value were disregarded. The data ranges
from years 1962 to 2018. (ii) A Wikipedia
unweighted bipartite network from 2026, where the top
1000\cite{Wikitop1000} (999
available) editors were used with articles edited by at least 10 of
them which resulted in 2,573,252 articles.

In the trade network the measure converges in 32 out of the 57 available years, which means
that for 44\% of the data, the self-consistent measure fails to converge.  Let
us examine two years where convergence does not occur: 1965 and 2015.
The results are presented in the top two rows of
Fig.~\ref{Fig:samplenoconv} which displays two different scenarios. In
1965 (top row), only a few products exhibit convergence for the
measure $C$, whereas in 2015, measure $D$ converges for only a single
country, but $C$ converges for most products.  However, a common
feature in both examples is that the data split into two groups: the
measure approaches zero in one group and remains constant ($\sim
\mathcal{O}(1)$) for the other. This suggests that the situation might
be similar to our initial $2\times 2$ Z-shaped example network
(Fig.~\ref{Fig:22}), where the diagonal link connects the two surviving (finite)
measures, while the measures tending to zero are unconnected.  The
Z-shaped network can obviously be mirrored due to the symmetry of the
measures. 

The measure also does not converge for the Wikipedia data which is
shown in the bottom panel of Fig.~\ref{Fig:samplenoconv}, where only a
subset of the available measures are shown due to the large number of
nodes. We observe a similar pattern with two branches, one decreasing
exponentially to zero, the other remaining constant but only due to
the normalization.  In this case however both branches are more
numerous than in the previous example due to the larger minimal degree
of the network.

We have thus shown that this self-consistent measure may have
convergence issues on any scale, ranging from $2\times2$ to
$10^3\times 10^7$ nodes.

\section{Convergence Criterion}

\subsection{General derivation}

Without loss of generality, we will assume that in the bipartite
network nodes in set $U$ are countries, and nodes in set $V$ are
products.  We assume that our network $N^o=\{U^o,V^o,E^o\}$ is a
connected but not fully connected bipartite network.  Let $N_D$ be the
cardinality of $U^o$ (number of countries) and $N_C$ the cardinality
of $V^o$ (number of products).  The suffix $o$ denotes the original
network.  The subscript refers to the measure associated with the
nodes to avoid confusion with the letter $C$ which could then mean
both country and complexity.

We
group all nodes in $U^o$ and $V^o$ into a condensed bipartite network
with two multi-nodes in each of the two node sets, $U=\{u_0,u_1\}$ and
$V=\{v_0,v_1\}$, such that the adjacency matrix of the condensed
network is a $2\times2$ matrix with one zero in one of the
off-diagonal elements.  This configuration creates a Z-shaped or a
flipped Z-shaped graph, as illustrated in Fig.~\ref{Fig:22}.  There
are numerous ways to create such a grouping, which will be discussed
later.    
We assign the $D$ measure to countries and the $C$ measure to
products.  The number of countries in multi-node $e\in\{0,1\}$ is
$n_{De}$, and the number of products in multi-node $\alpha\in\{0,1\}$
is $n_{C\alpha}$.
The weight matrix of the condensed network is
\begin{equation}
W=\begin{pmatrix}
W_{00}&0\cr W_{10}&W_{11} \end{pmatrix},
\end{equation}
where the weights are considered as the average weight between the nodes of the respective groups.  The condensed network is illustrated in Fig.~\ref{Fig:22}.

Our second assumption, a mean-field-like approximation, is that all products and countries within each multi-node group have the same complexity and scatteredness.
We denote these by $D_0$, $D_1$, $C_0$, $C_1$. In the iterative formula (\ref{Eq:origCD}), these mean-field assumptions simplify the sum to:
\begin{align}
D_0'&=\frac{1}{\mathcal{N}_D}\left(n_{C0}\frac{W_{00}}{C_0}\right)\quad
C_0'=\frac{1}{\mathcal{N}_C}\left(n_{D0}\frac{W_{00}}{D_0}
+n_{D1}\frac{W_{10}}{D_1}\right) \cr
D_1'&=\frac{1}{\mathcal{N}_D}\left(n_{C0}\frac{W_{10}}{C_0}
+n_{C1}\frac{W_{11}}{C_1}\right)\quad
C_1'=\frac{1}{\mathcal{N}_C}\left(n_{D1}\frac{W_{11}}{D_1}\right),
\end{align}
where the prime indicates the value in the next iteration and $\mathcal{N}_D$ and $\mathcal{N}_C$ are the normalization factors for $D$ and $C$, respectively. As shown below, these factors cancel and play no role in the criterion. Dividing the two-by-two equations yields the following iterative equations:
\begin{align}
\frac{D_1'}{D_0'}=&
\frac{W_{10}}{W_{00}}+
\frac{W_{11}}{W_{00}} \cdot \frac{n_{C1}}{n_{C0}} \cdot \frac{C_0}{C_1}\cr
\frac{C_0'}{C_1'}=&
\frac{W_{00}}{W_{11}} \cdot \frac{n_{D0}}{n_{D1}} \cdot \frac{D_1}{D_0}+
\frac{W_{10}}{W_{11}}
\end{align}
Note that the normalization factors cancel out. This is a simple iteration with $x=D_1/D_0$ and $y=C_0/C_1$, and the corresponding constants, leading to the equations:
\begin{align}
x'&=a_1y+b_1\cr
y'&=a_2x+b_2,
\end{align}
which converges if and only if $a_1a_2<1$~\cite{saad2003iterative}.  In our case, the condition is:
\begin{equation}\label{Eq:crit}
n_{D0}n_{C1}< n_{D1}n_{C0}
\end{equation}
%Another remark is that 
The criterion states that the product of the multiplicities along the diagonal link must be larger than that of the unconnected diagonal. This clarifies why our small example related to Fig.~\ref{Fig:22} failed to converge: instead of an inequality, we have an equality that does not satisfy the criterion.  It should be noted that %is worth noting
the criterion is independent of the network weights and depends solely on the network structure. This structural character connects our result to the earlier observation that the block structure of the biadjacency matrix governs which nodes are driven to zero~\cite{pugliese2016convergence}; the criterion~(\ref{Eq:crit}) makes this quantitative for the elementary $2\times2$ reduction. We stress that it characterizes the original, parameter-free map, and is thus complementary to the result that the \emph{regularized} ($\delta>0$) map possesses a unique fixed point by virtue of its convex potential~\cite{bellina2026cost}.%Another remark is that the criterion states that the product of the multiplicities along the diagonal link must be larger than that of the unconnected diagonal.

It is worth examining how this mean-field condition relates to the convergence of the full network, and in what sense it is exact. At the coarse-grained level the criterion is \emph{exact}: for each grouping the reduced map converges if and only if $a_1a_2<1$, so requiring~(\ref{Eq:crit}) for every grouping is both necessary and sufficient for the convergence of the mean-field dynamics. The only approximation is the replacement of the heterogeneous measures within a group by their average, and this approximation is \emph{one-sided}. Indeed, by the inequality between the harmonic mean $H$ and the arithmetic mean $\bar{x}$,
\begin{equation}
    H \equiv \frac{n}{\sum_i^n \frac{1}{x_i}} \le \bar{x} = \frac{\sum_i^n x_i}{n},
\end{equation}
the next-iteration value of a measure that tends to zero satisfies
\begin{equation}
    D_0'^{\rm (real)} \equiv \sum_i^{n_{C0}} \frac{W_{00}}{C_0^i} \ge n_{C0} \frac{W_{00}}{\bar{C}_0} = D_0',
\end{equation}
that is, the true update keeps the vanishing measure at least as large as the mean-field reduction does. The reduced $2\times2$ dynamics therefore drives the measures to zero \emph{at least as fast} as the true iteration: the reduction is a conservative proxy that can only over-estimate the tendency to non-convergence, never under-estimate it. As a consequence, for the original network the criterion is a rigorous \emph{sufficient} condition---if~(\ref{Eq:crit}) holds for all groupings, the network is guaranteed to converge---whereas the converse can fail only in the harmless direction: a grouping may violate the criterion while the true network still converges (a false alarm), but the criterion can never certify a non-convergent network as convergent. In our data we observed no such discrepancy: every non-convergent case exhibited a violating grouping involving only a few low-degree nodes, and no network flagged in this way subsequently converged. A distinct, purely computational limitation is that the number of possible groupings grows exponentially, so in practice the test is applied only to the groupings one can enumerate; we return to this point below.
% (TS: framing revised -- exact (iff) at mean-field level; one-sided/conservative error => rigorous SUFFICIENT for the real network, converse fails only as a false alarm. Please re-check.)

Let us verify if the examples in Fig.~\ref{Fig:samplenoconv} violate the above criterion.  In 1965, there were 152 countries and 690 products.  There is a country ($n_{D0}{=}1$) that exports only two products ($n_{C0}{=}2$), which would imply:
$
n_{D0}n_{C1}=1\cdot (690-2)=688 \not< n_{D1}n_{C0}=(152-1)\cdot2=302.
$
There is another country with 4 export products. If we group these two countries together ($n_{D0}{=}2$), they collectively export only 6 products ($n_{C0}{=}6$), to yield:
$
n_{D0}n_{C1}=2\cdot (690-6)=1368 \not< n_{D1}n_{C0}=(152-2)\cdot6=900.
$
The above two countries are the ones colored blue in the top left hand panel of Fig.~\ref{Fig:samplenoconv}, where it clearly seen that
the associated $D$ values do not converge. The two products exported by the first country are also red, the other four are blue but can be seen above the rest in blue in the top left panel of Fig.~\ref{Fig:samplenoconv}.

In 2015 (with 143 countries and 761 products), a single country exported $n_{C0}=6$ products ($n_{D0}=1$).  This corresponds to the flipped Z-shaped case, for which the criterion reads as follows
$
n_{D0}n_{C1}=1\cdot (761-6)=755 \not> n_{D1}n_{C0}=(143-1)\cdot6=852.
$
In the lower row of Fig.~\ref{Fig:samplenoconv} the country is colored red in the left hand panel, the corresponding products blue in the right panel.

In the general case, this criterion must be checked for all possible groupings.  However, as we have observed, in most cases, it fails with only a few items in one of the groups. Furthermore, we will present an algorithm that can easily make the data convergent. First, we will check some limit cases.

\subsection{Limit of large number of products}

We analyze another important case, assuming $N_C>N_D$, i.e., there are more products than countries.  We assign the country with the lowest degree ($k_{C,\mathrm{min}}$) to group 0, so $n_{D0}=1$. Naturally, $n_{C0}=k_{C,\mathrm{min}}$. The other multi-node cardinalities are: $n_{D1}=N_D-1$ and $n_{C1}=N_C-k_{C,\min}$.  Thus, the criterion becomes:
\begin{align}
1>&\frac{n_{D0}n_{C1}}{n_{D1}n_{C0}}=
\frac{1\cdot(N_C-k_{C,\min})}{k_{C,\min}(N_D-1)}\simeq
\frac{N_C}{N_Dk_{C,\min}}\cr
k_{C,\min} >& \frac{N_C}{N_D},
\end{align}
where in the last step, we assumed that $N_D\gg1$ and $N_C\gg k_{C,\mathrm{min}}$.  We have derived a simple criterion for convergence: the smallest degree must be greater than the ratio of products to countries.  The most straightforward solution to this problem is to remove countries with low degrees. 

In the opposite case, when we start with the product of the lowest degree, it always converges since, in general, there are more products than countries, so $N_D/N_C<1$.  However, in 
a more populous set of nodes, it can happen that many low-degree nodes are connected to a few nodes, which may also pose a problem.  Consider the extreme case where a country exports many products, but among them there are $S$ products that are not exported by anyone else, thus having a degree of one.  In this case, the bipartite network Z is flipped and $n_{C0}=S$, $n_{D0}=1$, $n_{C1}=N_C-S$, and $n_{D1}=N_D-1$.  The criterion is again that the product of the numbers along the diagonal of the Z must be larger than the product of the unconnected diagonal:
\begin{equation}
1>\frac{n_{C0}n_{D1}}{n_{C1}n_{D0}}=
\frac{S(N_D-1)}{N_C-S}.
\end{equation}
If $N_C\gg S$, the criterion simplifies to $S<N_C/N_D$.  This implies that if a node has more one-degree neighbors than the ratio $N_C/N_D$, the measures will not converge.  This situation is particularly relevant when $N_C\simeq N_D$, in which case $N_C/N_D\simeq1$ and even a single uniquely exported product ($S=1$) already violates the criterion, so that such products cannot be tolerated.  The solution to this problem can involve either grouping these products together or discarding them.

After removing countries with low degrees and products exported by a single country that violated the second relation, there was only one year, 1963, when the measure did not converge.  In this year, two countries together exported 9 products,  which violated the criterion, even though each country had a higher degree than required by the first inequality.

\subsection{Applicability of the criterion}

In the previous two sections, we presented two common scenarios that occur frequently, and the problem invariably involves low-degree nodes.  Unfortunately, the situation is not always so simple as the criterion can be violated by %very
complex scenarios.  For example, imagine a network with $N_C=1000$ products and $N_D=100$ countries. Suppose 100 of the products are exported by only 10 of the countries; grouping these 10 countries into multi-node $0$ ($n_{D0}=10$, $n_{D1}=90$) and those 100 products into the corresponding product multi-node ($n_{C0}=100$, $n_{C1}=900$), the criterion is violated because
\begin{equation}
    1\not>\frac{n_{D0}n_{C1}}{n_{D1}n_{C0}}=
    \frac{10\cdot 900}{90\cdot 100}=1.
\end{equation}
An exhaustive search over all possible groupings is combinatorially infeasible, since their number grows exponentially with the network size.  In the next section, we propose three algorithms to resolve the problem of non-convergence.

\section{Algorithms for convergence}

In this section, we detail three algorithms designed to induce convergence of
the measures.  All methods rely on the iterative calculation of self-consistent measures.  We check for convergence and, if the test fails, we can apply one of the following three methods: node removal, node merging, and regularization. Convergence is assessed from the evolution of the measures with the iteration number $n$. Because the measures can oscillate from one step to the next, a single-step difference is not a reliable indicator; instead we monitor the change of the measures over a window of several (typically ten) consecutive iterations and base the decision on this longer-term behavior, declaring convergence only when the variation across the window stays below a small threshold.  %number.

\subsection{Node removal}

The first proposed method is node removal, which has been empirically adopted in Wikipedia analysis~\cite{ogushi2021ecology}. The problem consistently lies with nodes whose measures approach zero.  We aim to identify the less numerous group whose measure decreases to zero. This can be achieved by comparing the mean and minimum values, with a larger difference indicating that the group tending to zero is less numerous.  One can remove either all nodes with decreasing values or successively remove the node with the smallest measure until convergence is achieved.  In case of our data, %in order
to achieve convergence, only countries with five or fewer export products were deleted, and goods exported by a single country or, on four occasions, by two countries were deleted.

Removing nodes modifies the network and may, in principle, affect the measures of the retained nodes, so the operation must be applied with care. In practice, however, the affected nodes are precisely the marginal, low-degree ones whose measures are already collapsing to zero; they carry little information, and their removal leaves the rest of the ranking essentially unchanged. In many cases the operation is also justified on substantive grounds. For the Wikipedia network, for instance, the problematic authors had relatively low impact on the remaining article set. By applying the rule that only those articles are kept which were edited by at least 10 editors, these editors lost 99\% of their contribution. Therefore excluding them from the study is actually justified.

\subsection{Node merging}

Removing nodes can have undesirable effects as, for example, important but specialized products can then be removed. A more gentle approach involves merging the problematic nodes. This is particularly useful for goods, since the classification already contains many near-duplicate categories that can be combined with little loss of meaning, for example \emph{Glass}, \emph{Glassware}, \emph{Miscellaneous Articles of Glass}, and \emph{Miscellaneous Glass}.  The algorithm is similar to the previous one: we identify the nodes whose measure decreases to zero in the less numerous group, and we combine the two nodes with the two smallest measures into one by summing the corresponding rows (or columns) of the weight matrix.  This process is repeated until convergence is reached.

As with node removal, merging alters the network structure, so in any concrete application the specific combinations should be inspected to ensure that genuinely distinct nodes are not conflated; the near-duplicate categories above illustrate cases where this is unproblematic. Where no such substantive justification is available, the regularization scheme discussed next offers an alternative that restores convergence while leaving the node set untouched.

\subsection{Regularization}

A third way to restore convergence, which leaves the node set untouched, is regularization. Adding a non-homogeneous term to the Fitness--Complexity map is not a new idea: it was introduced by Servedio \textit{et al.}~\cite{servedio2018stable}, whose non-homogeneous estimator augments each update with a constant that prevents the measures from collapsing to zero. In our notation it reads
\begin{equation}\label{Eq:NHEFC}
D_e^{(n)} = \delta + \sum_\alpha \frac{w_{e\alpha}}{C_\alpha^{(n-1)}},
\qquad
C_\alpha^{(n)} = \delta + \sum_e \frac{w_{e\alpha}}{D_e^{(n-1)}},
\end{equation}
with \emph{no} normalization step: the constant $\delta$ fixes the otherwise free overall scale and thereby removes the need to normalize. As $\delta\to0$ the original homogeneous metric is recovered up to a rescaling, so $\delta$ should be taken small in order to leave the ranking essentially unchanged~\cite{servedio2018stable}. The fixed point of this map is by now well understood: the Fitness--Complexity iteration is equivalent to the classical Sinkhorn--Knopp matrix-scaling algorithm~\cite{mazzilli2024equivalence}, and the regularized map can be obtained as the minimizer of a convex potential, which guarantees a unique fixed point and global convergence for broad classes of networks~\cite{bellina2026cost}.

Unlike node removal and merging, which retain the original scheme and modify the network, regularization replaces the dynamics: by construction the non-homogeneous map~(\ref{Eq:NHEFC}) is not subject to the non-convergence analyzed above, and in all the cases we examined it converged to a well-defined fixed point. A naive alternative that instead merely adds a constant to the \emph{normalized} scheme does not share this property---because the normalization rescales the constant away at every step, it neither guarantees convergence nor admits a small regularizer. For this reason we adopt the non-homogeneous form~(\ref{Eq:NHEFC}), with a single small $\delta$ applied uniformly to all datasets, as the regularization remedy in the comparison below.

\section{Results}

\begin{figure*}
\centering
\centerline{Node removal}
\includegraphics[width=.90\textwidth]{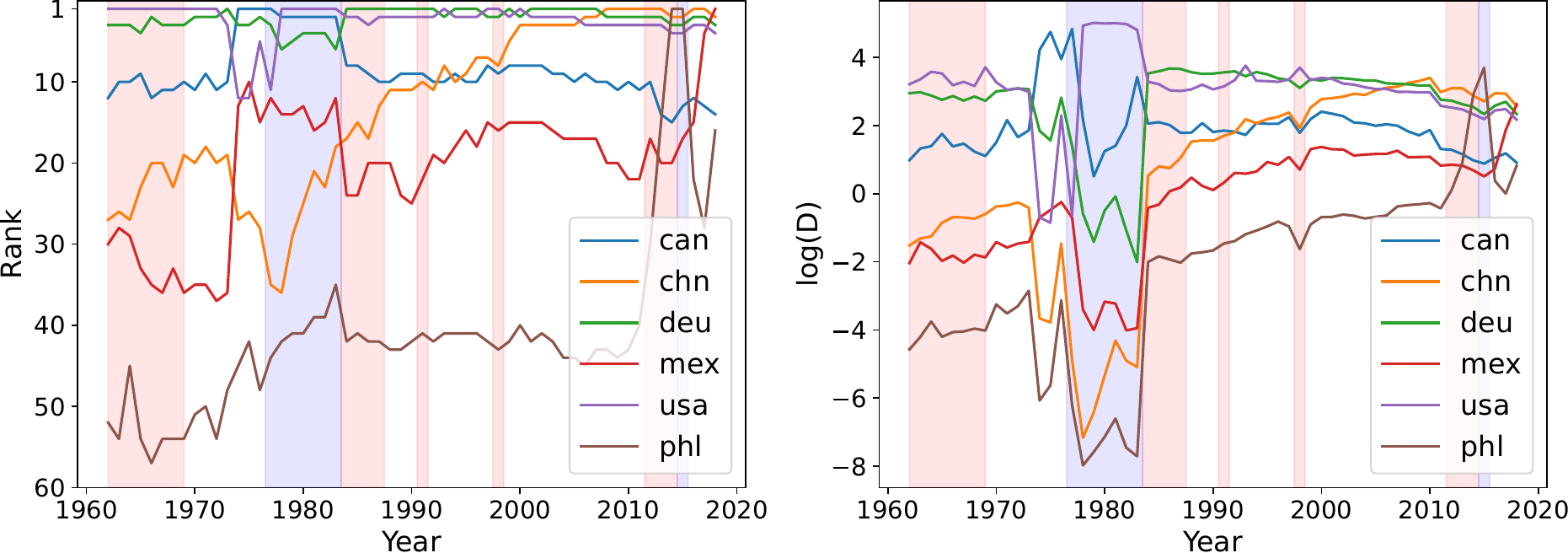}
\centerline{Node merging}
\includegraphics[width=.90\textwidth]{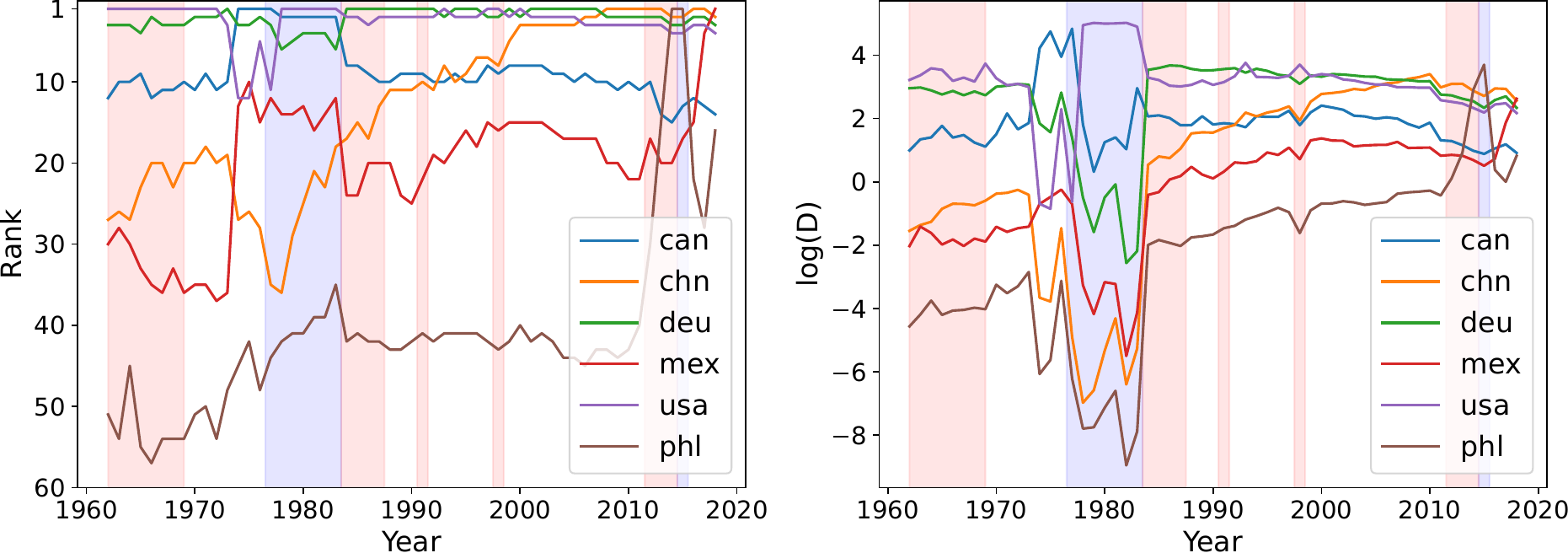}
\centerline{Regularization}
\includegraphics[width=.90\textwidth]{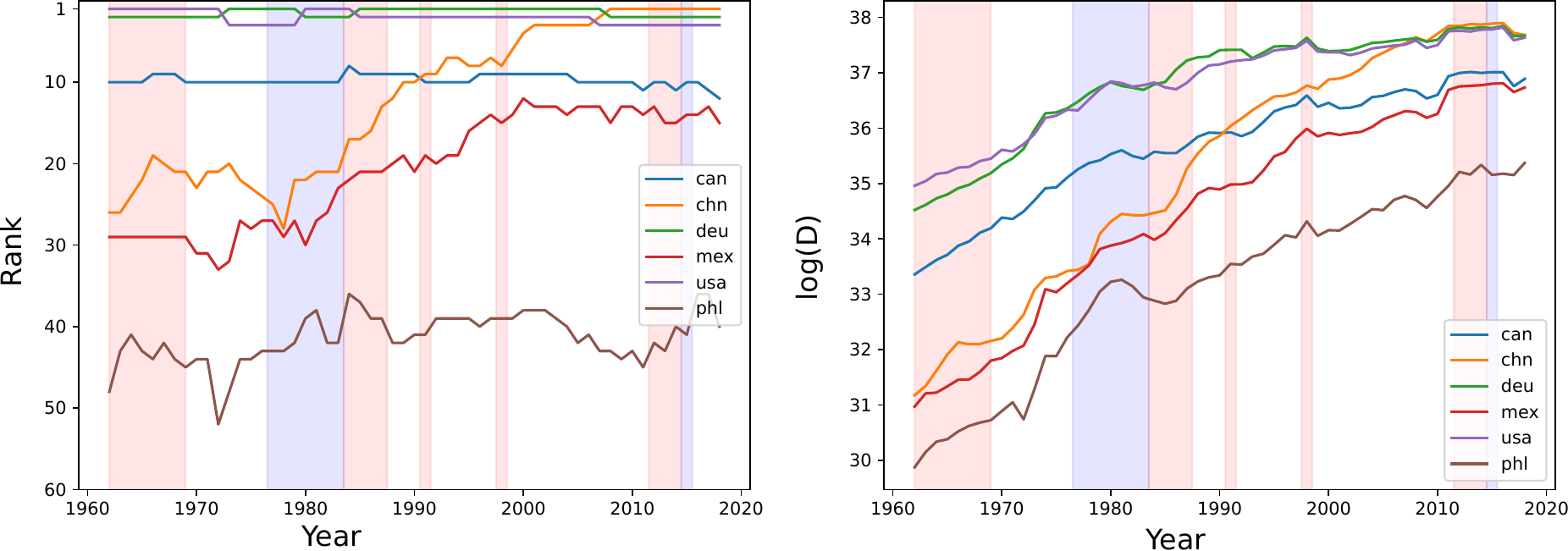}
\caption{The results of the three methods, from top to bottom: node removal, node merging, and regularization.  Only countries that consistently had the largest Fitness $D$ value are shown; Canada (can), China (chn), Germany (deu), Mexico (mex), USA (usa), and Philippines (phi). Left: ranking as a function of years.  Right: the value of $\log(D)$ as a function of the years.  The graphs are shaded red if countries caused non-convergence, and blue regions indicate goods-related convergence issues.}
\label{Fig:3comp}
\end{figure*}

Before examining the results, it is important to note that for our empirical data, in all cases the algorithm's behavior coincided with the theoretical predictions.  If the network did not converge, we could always find a corresponding $2\times2$ network that violated our criterion.  Due to computational limitations, we did not formally prove the opposite case, but we verified that for the simplest scenarios with $n_{C0}=1,2$ and $n_{D0}=1,2$, the criterion was satisfied. 

In Fig.~\ref{Fig:3comp} %We compare 
the results of all three methods %in Fig.~\ref{Fig:3comp}
, where regions with convergence issues are shaded.  Naturally, the curves are identical outside these regions. Although the curves appear similar, there are interesting features.  A mysterious behavior is observed around 1980, but this disturbance began earlier in the 1970s, coinciding with a significant increase in global trade. Nevertheless, the Fitness $D$ values for %in 
this period differ substantially from those before and after. This indicates that, near the stability threshold, convergence becomes very slow, so that stopping the iteration as soon as the network appears stable can leave residual artifacts. This motivates the \textit{parallel} removal introduced below, which eliminates all problematic nodes in a single step, moving the network away from the threshold and restoring fast, clean convergence.

Another interesting feature is the Philippines' top ranking in 2014-2015.  How could a country that typically ranked around 40th in the Fitness $D$ measure suddenly become first, then drop back 20 places?  The data of Philippines do not suggest any sudden trading burst, as volume and degree change continuously.  The reason is that in 2015, the network is close to instability in both senses detailed above: there is a country (Angola) with a degree just a fraction larger than required, and the Philippines has just a fraction fewer unique export products than required for stability.  In fact, the network was made convergent by removing a unique export product (or merging two unique export products) from the Philippines.  After this step, the measure converges, but very slowly, and will be dominated by these features.  Thus, the value of Fitness $D$ in the Philippines will not become infinite but will be large. 

\begin{figure}
\centering
\includegraphics[width=1.05\columnwidth]{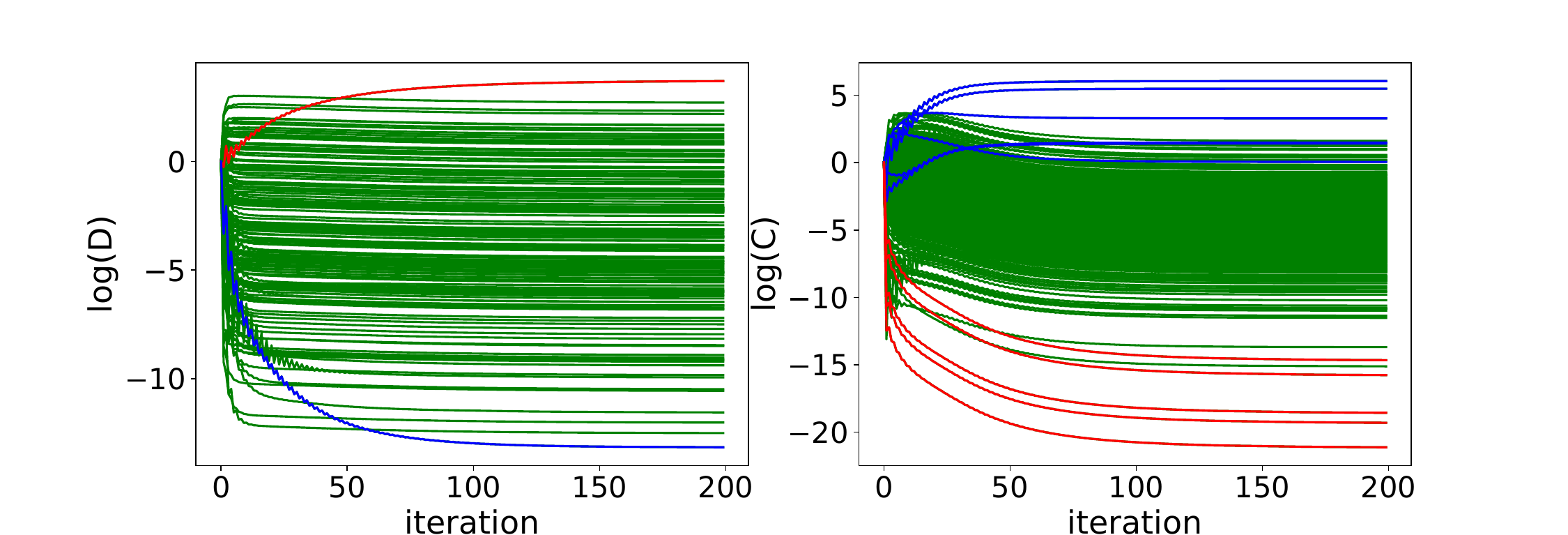}
\caption{The evolution of Fitness measure $D$ (left) and Complexity measure $C$ (right) through the iteration for the year 2015. On the left, the blue curve is Angola, the red one the Philippines, and green the rest of the world; on the right, the red curves are the complexity $C$ of the products exported only by the Philippines, the blue ones those exported by Angola, and the green ones the remaining products.}
\label{Fig:rem_2015}
\end{figure}

In Fig.~\ref{Fig:rem_2015} we show the evolution of the measures through the iteration; Angola (blue) and the Philippines (red) are clearly distinguishable in the left-hand plot. They converge, but only very slowly. The underlying problem is the same as before: many products are exported solely by the Philippines. Whether this reflects a database artifact or reality is beyond the scope of this paper, but removing all these products at once resolves the non-convergence, and the Philippines returns to its usual place. We call this variant of the removal process \textit{parallel}: all non-converging nodes---products or countries whose measure decreases to zero---are removed in a single step, in contrast to the \textit{serial} process that removes them one at a time.

\begin{figure*}
\centering
\includegraphics[width=.95\textwidth]{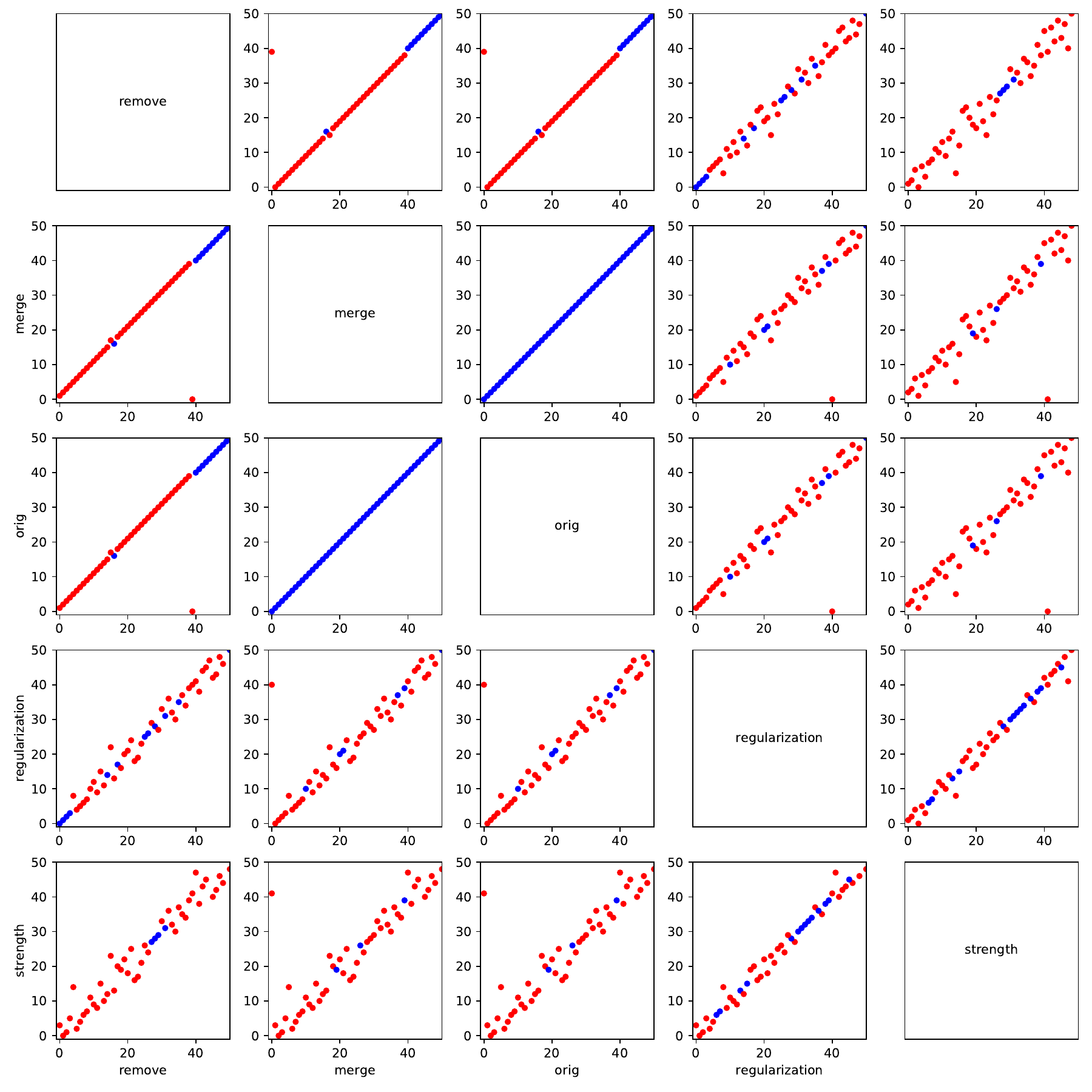}
\caption{Comparison of the top-50 country rankings for the year 2015 using the methods parallel remove, serial merge, orig, regularization, and strength (defined in the text). Blue points indicate matching ranks.}
\label{Fig:rank}
\end{figure*}

As a final comparison we visualize the rankings produced by the different methods. For each method we computed the country measure and ranked the countries in decreasing order; Fig.~\ref{Fig:rank} plots these rankings against each other for the year 2015. The methods are:
\begin{itemize}
\item \textbf{parallel remove}: all products whose measure would go to zero were removed at once ($6$ products in 2015).
\item \textbf{serial merge}: the product with the lowest scatteredness was merged with the one with the second lowest; the $C$--$D$ measures converged after a single such step.
\item \textbf{orig}: the original homogeneous scheme stopped at rank convergence, i.e.\ when the rank order stabilized even though the values had not.
\item \textbf{regularization}: the non-homogeneous scheme of Servedio \textit{et al.}~\cite{servedio2018stable} with a small constant $\delta=10^{-5}$.
\item \textbf{strength}: countries ranked by their node strength, i.e.\ the total weight (export volume) of their links.
\end{itemize}
We do not show \textit{serial remove} separately, as it was identical to \textit{serial merge} for the top 50 countries.

The plots show that, in 2015, the original homogeneous scheme (orig, stopped at rank convergence) places the Philippines at the top, whereas every convergence-restoring method places China there instead. At the same time, node merging leaves the order of the most important countries essentially unchanged relative to original rank convergence method; across other years only an occasional swap of neighboring ranks occurs. This justifies our remedies: they preserve the ordering of the original homogeneous scheme while additionally yielding a value-converged result, so that the actual measure values---not only the ranks---can be used.

Removing the $6$ products unique to the Philippines moves the network well away from the stability limit, so the parallel variant converges quickly and cleanly, rather than through the slow transient of the serial process. The non-convergence is thereby traced to a handful of loosely coupled nodes---here the products exported only by the Philippines---whose removal alone restores a well-defined ranking.

\begin{figure}
\centering
\includegraphics[width=.95\columnwidth]{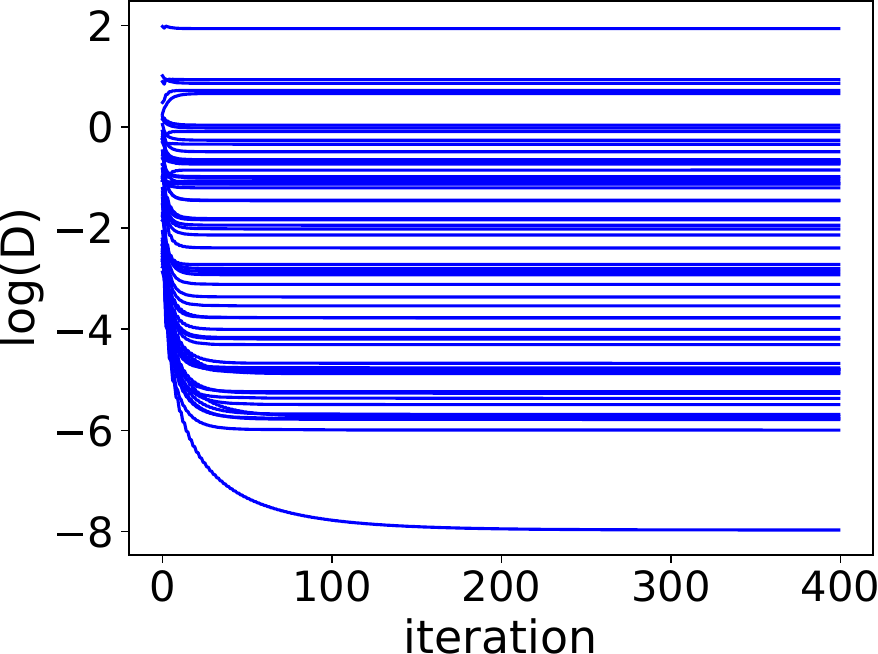}
\includegraphics[width=.95\columnwidth]{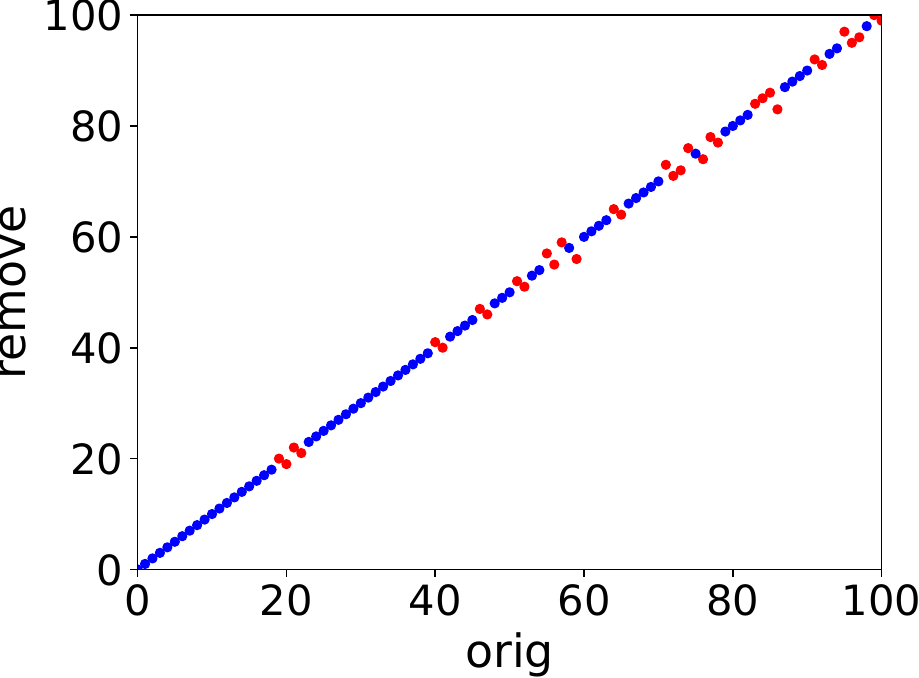}
\caption{Convergence remedy on the Wikipedia editor--article network. Left: evolution of the measure through the iteration after removing the low-degree editors; the network converges, with a single editor remaining close to the stability boundary---its criterion satisfied only by about $5\%$---and detaching from the rest. Right: the editor ranking of the corrected network plotted against that of the uncorrected one, showing only marginal differences.}
\label{Fig:Wiki}
\end{figure}

We have also analyzed a Wikipedia editor--article bipartite network, in which editors play the role of countries and articles that of products. As shown in Fig.~\ref{Fig:samplenoconv}, the original measure does not converge; consistently with our criterion, the smallest editor degree is below the required ratio $N_A/N_E$ (here we are in the limit $N_A\gg N_E$ of many more articles than editors). Removing all editors whose degree falls below this threshold (23 editors) restores convergence. One editor nevertheless remains close to the stability boundary---its criterion is satisfied only by a margin of about $5\%$---and accordingly detaches from the rest in Fig.~\ref{Fig:Wiki}, the direct analogue of the Philippines in the trade network. The Wikipedia network thus confirms both our convergence criterion and the proposed remedies in a setting structurally different from trade. Moreover, the editor ranking of the corrected network differs only marginally from that of the uncorrected one, so---as in the trade case---the remedy restores convergence without distorting the ranking.

Returning to the trade network, we compared the homogeneous and non-homogeneous rankings with the ranking by node strength, the natural zeroth-order baseline. The non-homogeneous measure turns out to be more closely related to node strength than the homogeneous one: across the years examined, its Pearson correlation with the strength ranking is in general about twice as close to unity (the perfect straight line) as that of the homogeneous measure. We report this as an observation; what it implies for the relative merits of the two schemes is beyond our scope.

We stress that it is not the aim of this work to determine which scheme or remedy yields the ``best'' ranking; the methods differ in detail, and the appropriate choice depends on the application. What matters here is that the proposed convergence-restoring methods do not significantly alter the ranking of the original scheme, while making its values well defined.

\section{Summary}

In summary, we have demonstrated that the self-consistent measures of Fitness and Complexity, or Scatteredness and Complexity, respectively, do not always converge for bipartite graphs.  Convergence is independent of the network weights and depends uniquely on the network structure.  We established a criterion for the $2\times2$ graphs and showed that using
this as a mean-field representation for large complex networks
provides an excellent estimator for the convergence of the original network. Unfortunately, testing all possible $2\times2$ representations of the original network is computationally infeasible, but in our practical examples, we never encountered problematic cases with $n_{C0}>2$ or $n_{D0}>2$.

To address instances of non-convergence, we discussed and compared three procedures aimed at restoring convergence: removal of problematic nodes, merging of nodes, and regularization. The first two operate on the original scheme, while regularization follows the non-homogeneous construction of Servedio \textit{et al.}~\cite{servedio2018stable}. These approaches resolve the convergence failures in all of our examples and agree on the top of the ranking, although they differ in detail further down; determining which is preferable is beyond our scope. We further found that near the stability threshold convergence is slow, so that stopping the iteration as soon as the ranking appears stable can leave a spurious result (e.g.\ the Philippines ranked first in 2015); applying the remedies---in particular removing the few loosely coupled nodes responsible---restores a value-converged ranking in which this artifact disappears. In this way our criterion and remedies render the scheme usable, ensuring a value-converged result rather than one that depends on the stopping rule. We illustrated this on two structurally different empirical systems, the international-trade network and the Wikipedia editor--article network, on which the criterion and the remedies behave consistently. Our study thus provides both a theoretical framework and practical tools to ensure the reliable use of self-consistent measures across diverse bipartite systems.

\section*{Acknowledgments}

JT acknowledges the support from Advanced Grant 149429 from the
Hungarian National Research, Development and Innovation Office, and Grant TKP2021-NVA-02 of the National Research, Development and
Innovation Fund of Hungary. This work was partially supported by JSPS KAKENHI grant numbers JP21K19826 and JP23K03256 to TS and FO.
KK and JK acknowledge support from EU HORIZON 2020 INFRAIA-1-2014-2015 program project (SoBigData) No. 654024 and INFRAIA-2019-1 (SoBigData++) No. 871042. KK acknowledges NordForsk Programme for Interdisciplinary Research project “The Network Dynamics of Ethnic Integration”. JK acknowledges partial support from EU grant ERC Synergy Grant No 810115 DYNASNET. JT, TS, FO, and JK thank Aalto University for hospitality. We are especially grateful to Cesar Hidalgo for the trade data. The authors used Claude (Anthropic, Claude Opus 4.8) to assist with literature search, language editing, and manuscript revision. The authors directed the tool, and reviewed and verified all of its output, taking full responsibility for the work.

\bibliography{lit}
\end{document}